\let\Xdocument\document

\documentclass{iau}

\usepackage{amsmath}
\usepackage{gensymb}
\usepackage{graphicx}
\usepackage{xcolor}
\usepackage[version=4]{mhchem}

\newcommand\farcs{.\!^{\prime\prime}}

\let\document\Xdocument

\begin{document}

\lefttitle{Catherine Walsh}
\righttitle{C/O as a diagnostic of planet formation}

\jnlPage{1}{7}
\jnlDoiYr{2024}
\doival{10.1017/xxxxx}

\aopheadtitle{Proceedings IAU Symposium}
\editors{E.~A.~Bergin, P.~Caselli, J.~K.~J\o rgensen}

\title{Linking planet formation to exoplanet characteristics: C/O as a diagnostic of planet formation}

\author{Catherine Walsh}
\affiliation{School of Physics and Astronomy, University of Leeds, Leeds, UK, LS25 1HZ}

\begin{abstract}
Gas-giant exoplanets are test cases for theories of planet formation as their atmospheres are proposed to carry signatures of their formation within the protoplanetary disk. The metallicity and C/O are key diagnostics, allowing to distinguish formation location within the disk (e.g., relative to snowlines), and mechanism (e.g., core accretion versus gravitational instability). We can now probe the composition of the planet-forming regions of disks, and that in gas-giant exoplanets, to scrutinise these theories and diagnostics. So far, ALMA has revealed that the outer disk regions are typically metal-depleted and O-poor, whereas JWST is showing that the inner disk regions around Sun-like stars are mostly O-rich. Further, JWST is showing that most transiting gas-giant planets are typically metal-enriched and O-rich, consistent with formation at/within the water snowline and pollution by icy bodies. There is emerging an arguably ``conventional" picture of gas-giant planet formation for transiting planets, to be confirmed, of course, with future data.
\end{abstract}


\maketitle

\section{Introduction}
\label{introduction}

At the time of writing (early 2024), we have detected more than 5,600 planets in more than 4,100 planetary systems orbiting other stars\footnote{\url{https://exoplanet.eu/catalog/}}. 
We have moved from the highly prolific era of exoplanet \emph{detection} and into the era of exoplanet \emph{characterisation}. 
Thanks to enormous past efforts with the \emph{Hubble} Space Telescope (HST) \citep[e.g.,][]{Sing16}, and high-resolution ground-based spectrometers such as VLT/CRIRES \citep[e.g.,][]{Birkby17}, we have been able to compile a census of molecules present in the some of the most extreme exoplanets that we have detected to date, the so-called \emph{hot Jupiters} (see recent reviews on this class of planets by \citealt{Dawson18} and \citealt{Fortney21}).
These are named for their mass (from $\approx 0.3 - 13~\mathrm{M_{J}}$) and 
proximity to their host star (with orbital periods $\lesssim 10$~days), which also makes them prime targets for atmospheric characterisation via transmission and emission spectroscopy.  
Because they are gas giants, they are thought to have better retained a chemical fingerprint of their formation mechanism and history than lower-mass planets; however, as we will see later, this picture has become more unclear as we learn more and more about the process of planet formation itself.
We also now have a reasonably-sized population of directly-imaged gas-giant planets which can be characterised by direct emission spectroscopy \citep[see the recent review by][]{Currie23}.  
It should be noted that this planet population tends to be more massive and younger than the population of hot Jupiters detected via the transit method, and they tend to be discovered around higher-mass (i.e., A-type) stars.  
These properties aid their detection via direct imaging and characterisation via emission spectroscopy as they still possess significant luminosity from their formation.

Early efforts with HST showed that many hot Jupiters possess \ce{H2O} in their atmospheres, with additional reported detections of \ce{TiO}, \ce{AlO}, \ce{CO}, \ce{NH3}, \ce{CH4}, and \ce{HCN} in a few (not always the same) objects \citep[for an overview, see Table 1 in][]{Madhusudhan19}. 
The rise of retrieval tools of various levels of complexity and sophistication applied to exoplanet spectra now facilitates constraints, not only on the atmospheric composition, but also on the atmospheric metallicity, i.e., the ratio of heavy elements to lighter elements \citep[e.g.,][]{Madhusudhan18}. 
This measurement alone has been proposed to constrain the formation mechanism of the planet: a metallicity similar to the host star could indicate potential formation via direct fragmentation and collapse, whereas a metallicity higher than that of the star, could indicate formation via core accretion, similar to that proposed for the solar system gas- and ice-giants \citep[e.g.,][]{Helled14}. 
The solar system gas- and ice-giants have a metallicity that both increases with increasing semi-major axis, and with decreasing planetary mass, and has led to the theory that planet formation via  core accretion generates a population of planets with this intrinsic mass-metallicity relation \citep[e.g.,][]{Thorngren16}. 
Further it has been proposed that, if formed via core accretion, then the formation location of gas-giant planets relative to key snowlines in the midplane of the protoplanetary disk will leave a chemical fingerprint in both the C/O ratio and metallicity in the planetary atmosphere \citep[][]{Oberg11,Madhusudhan12}. 

The era of JWST has increased the number of planets with undisputed molecular features in addition to \ce{H2O}, also pushing the regime of exoplanet characterisation into and beyond the sub-Neptune mass range \citep[e.g.,][]{Madhusudhan23,Zieba23}.  
This now allows strong tests of the proposed mass-metallicity relationship and whether key snowlines play a role in setting (at least the initial) composition of gas-giant atmospheres. 
Analyses of both emission and transmission spectra of the hot gas giants published thus far suggest a mixed picture with reports of both super-stellar \citep[e.g., HD~209458~b;][]{Xue24} and sub-stellar metallicities \citep[e.g., WASP-77A~b;][]{August23}.
The presence of disequilibrium chemistry in the atmosphere (i.e., the effects of photochemistry and strong mixing) can also complicate the relationship between the metallicity inferred from the observations and the global metallicity of the planetary atmosphere \citep[e.g.,][]{Baxter21,Kawashima21}. 
Nonetheless, we are now at the point of being able to stress-test many of these proposed theories of what the atmospheric composition can tell us about the formation mechanism and pathway of this population of planets. 
We also now have in hand more than a decade of observations of protoplanetary disks with the Atacama Large Millimeter/submillimeter Array (ALMA) giving us unique insight into the composition of the gas accreted by planets at the epoch of planet formation.   
We are now moving into a new era of comparative studies between measurement of the composition and metallicity of the atmospheres of exoplanets with direct observations of the planet-forming regions of protoplanetary disks.

In this short (non-exhaustive) review, I will cover some of what we have learned about the composition of the planet-forming regions of protoplanetary disks with ALMA, focusing on constraints on the C/O ratio and the metallicity of the planet-building gas.
I will also highlight some of the recent results with JWST that give us a different, yet complementary, view of the planet-forming region.  I'll contrast this with quantitative details on what we have learned about the C/O ratio and metallicity in the population of hot Jupiters for which such constraints now exist.  
Throughout I will mention some of the processes occurring in protoplanetary disks that may complicate and confuse the simple relation between formation location/mechanism and the atmospheric metallicity.  
I'll finish with some final remarks and thoughts for the future.
For more comprehensive overviews on the chemistry of planet formation and of planet-forming regions, see the recent reviews by \citet{Oberg21a}, \citet{vanDishoeck21}, and \citet{Oberg23}.

\section{What have we learned about the composition of the planet-forming regions of protoplanetary disks?}
\label{PPDs}

\subsection{Insights from a decade of ALMA studies of protoplanetary disks at moderate resolution}
\label{PPDsALMAmod}
It has long been known that protoplanetary disks around nearby young stars are hosts to molecular disks in Keplerian rotation about their host star revealed by observations at (sub)mm wavelengths \citep[e.g.,][]{Dutrey14}.  
Early studies with the available single-dish and interferometeric facilities (e.g., IRAM 30-m, JCMT, PdPI, and SMA) revealed that protoplanetary disks are host to small, simple, organic molecules, also commonly observed towards star-forming clouds 
\citep[e.g., \ce{CO}, \ce{HCO+}, \ce{N2H+}, \ce{HCN}, \ce{CN}, \ce{CS}, and \ce{H2CO};][]{Kastner97,Thi04,Dutrey07,Oberg10,Dutrey11,Chapillon12}.  
However these observations were spatially unresolved in the case of single-dish observations, and only moderately resolved (and only for the largest disks) with the previous generation of interferometers which had typical beam sizes of a few arcseconds. 

ALMA is the first telescope to allow a zoomed-in view of the composition of planet-building material at high resolution ($\sim 0\farcs1$). Early observations at moderate resolution with ALMA ($0\farcs5 - 1\farcs0$) revealed almost instantly that the molecular emission from protoplanetary disks is not always smooth, and in the same manner as the dust emission, can possess significant sub-structure. 
For example, rings of \ce{N2H+}, \ce{CN}, and \ce{C2H} emission were seen in our nearest protoplanetary disk, TW Hya, which is also conveniently in a face-on orientation \citep[][]{Qi13,Teague16,Bergin16}.  
Observations such as these were our first hint that there may be significant radial chemical gradients in disk composition. 
Particularly intriguing is that the emission rings listed above all arise from different chemical and physical effects: \ce{N2H+} emission is a known tracer of the location of the CO snowline \citep[due to the gas-phase reaction, \ce{N2H+ + CO \longrightarrow HCO+ + N2}; e.g.,][]{vantHoff17}, \ce{CN} emission rings can arise from the variation in strength and shape of UV radiation as it propagates through the disk atmosphere \citep[e.g.,][]{Cazzoletti18}, and \ce{C2H} ringed emission (and that from hydrocarbons more generally) is an indicator of the transition from oxygen-rich to carbon-rich gas likely due to dust evolution \citep[e.g.,][]{vanderMarel21}. 
Moderate resolution observations also revealed chemical emission rings in \ce{DCO+} \citep[also related to the CO snowline location via the reaction, \ce{CO + H2D+ \longrightarrow DCO+ + H2};][]{Oberg15,Carney18}, \ce{HC3N} \citep[related also to the transition to carbon-rich gas;][]{Cleeves21}, and CS \citep[thought to be tracing localised gas depletion;][]{Teague17}.

These early, moderate-resolution, observations with ALMA also revealed some intriguing trends across the disk mass range. 
Observations of emission from CO and its rarer isotopologues (e.g., \ce{^{13}CO} and \ce{C^{18}O}) showed that disks around low-mass stars appear depleted in CO gas compared with the expected canonical abundance ratio of $\sim 10^{-4}$ with respect to \ce{H2} \citep[e.g.,][]{Ansdell16,Miotello19}.
The level of CO depletion was above that expected from already known CO depletion processes such as photodissociation (which is active in the disk atmosphere and in the outer disk) and freeze-out of CO as ice onto dust grains \citep[which is active in the disk midplane beyond the CO snowline at temperatures $\lesssim 20$~K; e.g.,][]{Williams14}. 
Depletion of CO gas complicates any derivation of gas mass from CO observations, as we break the common scaling factors used between dust and gas (dominated by \ce{H2}) and CO.  
Arguably, the determination of the available gas mass in a protoplanetary disk is one of the fundamental parameters important for planet formation: are these disks depleted in gas (i.e., they have a reduced gas-to-dust mass ratio) or are they depleted in CO (i.e., they have a reduced CO-to-\ce{H2} abundance ratio)? 

We have indications that the latter is the case and this comes from three lines of evidence: i) the detection of HD emission from several protoplanetary disks with the \emph{Herschel} Space Observatory which is a ``cleaner" tracer of disk gas mass as it is unaffected by chemistry \citep[e.g.,][]{Bergin13,McClure16,Kama20}, ii) that this apparent CO or gas depletion is not as significant in warm disks around higher-mass stars \citep[although it is seen in ``cold" disks around higher-mass stars; e.g.,][]{Kama16a}, and iii) protoplanetary disks with evidence of potential CO depletion have outer disks composed of carbon-rich gas revealed by bright emission from hydrocarbons \citep[][]{Miotello19}. 
This suggests that this is an effect influenced by the temperature and chemistry of the disk, rather than, e.g., related to the dispersal state of the gaseous disk.  
There is statistical evidence that disks around higher-mass stars disperse at a faster rate than those around lower-mass stars \citep[e.g.,][]{Ribas15}, which is in contradiction to the gas depletion scenario. 

Quite extreme CO depletion factors, up to $\sim 100$ \citep[e.g.,][]{Zhang17}, have been reported, which begs the question: what has happened to the CO? 
We require a mechanism that can remove CO from the gas-phase, and in the upper regions of the disk from which the bulk of the CO emission arises.
This is a challenge because CO is a hypervolatile molecule, only freezing out at low temperatures ($\lesssim 20$~K). 
Further, CO is also a rather robust molecule due to its strong intramolecular bond (11~eV) and its ability to self-shield from photodissociation. 
This enables CO to survive in the gas phase to higher scale heights in the disk than other molecules.
Depletion of CO can be done through a combination of chemistry and dust evolution within the disk \citep[e.g.,][]{Bosman18,Schwarz18,Krijt20}.  
At moderate temperatures, it is possible for CO to have a non-negligible residence time on the dust grains and to undergo surface reactions to form less volatile species such as \ce{CO2} and \ce{CH3OH} that will remain frozen out to higher temperatures \citep[e.g.,][]{Eistrup18,vanTerwisscha22}.  
However, it is not sufficient to only chemically convert the CO to a less volatile form, also needed is the gradual removal of oxygen from the gas, necessary to also explain the transition to carbon-rich gas seen in the outer regions of several protoplanetary disks.
This can done by evolution of the dust grains in the disk, which, in the outer regions, will be coated in water-rich ices \cite[e.g.,][]{Du15,Kama16b}.

That dust grains evolve in their size and density distributions in protoplanetary disks is well established \citep[see the recent review by][]{Birnstiel24}.  
The processes that dominate in the outer disk are growth, followed by gravitational settling to the midplane, and radial drift inwards towards the star.  
We see strong evidence of dust settling and drift of large grains in high resolution of thermal emission from dust grains at mm wavelengths \citep[see the recent review by][]{Andrews20}.
Often, images show that the mm dust disk extends out only to a fraction of the gas disk (an indicator of radial drift of dust relative to the gas see, e.g., \citealt{Birnstiel14} and \citealt{Facchini17}), and a comparison of the scale height of the emission between the mm dust and the gas (traced in CO) also indicates that the dust grains are well settled into a dust-rich sub-disk confined to the midplane (e.g., \citealt{Pinte16} and \citealt{Doi21}). 
This can also be seen directly in images of dust thermal emission from close-to edge-on systems \citep[e.g.,][]{Villenave20}.

So we can currently account for CO depletion in the outer regions of protoplanetary disks though a combination of dust evolution and chemical evolution that has sequestered or trapped oxygen (that would otherwise be available to reform CO in the gas phase) in the ice reservoir that has followed the path of the large dust grains and is now confined to the inner disk midplane.  
The radial drift of these ice-coated dust grains has another effect, in that it can efficiently transport frozen material from the outer disk to the inner disk, thus potentially enriching the inner disk midplane gas \citep[e.g.,][]{Booth19}.
So why would such processes matter for planet formation?  
Firstly, the concentration of solid matter in the midplane likely helps in the formation of planetary cores (assuming formation via core accretion), and secondly, and we will come back to this later, this is evidence that protoplanetary disks possess potentially extreme vertical \emph{and} radial gradients in their gas composition and metallicity.  
Hence, moderate-resolution studies with ALMA have shown us that the composition of the gas in protoplanetary disks is not homogeneous across the disk, is influenced by both chemistry and dust processes, that in turn, will influence the composition of gas-giant planets that are forming therein.
This is one of several mechanisms that can break the assumed link between gas-giant composition or metallicity with formation location in the disk with respect to the positions of key snowlines.
Further this phenomenon of mm dust (or pebble) drift, can also influence the composition of the inner disk: disks in which icy pebbles have undergone efficient radial transport are expected to exhibit a more oxygen (i.e., water) rich inner disk within the water snow line than those that have not \citep[e.g.,][]{Kalyaan23}.  
Pebble drift can also increase the metallicity of the gas in the inner regions with respect to the outer regions, and to the stellar metallicity \citep[e.g.,][]{Zhang20}.
Enrichment in carbon relative to oxygen is also possible, as icy pebbles cross key snowlines of carbon-rich molecules such as \ce{CH4} \citep[e.g.,][]{Booth19}. 
However, as we will discuss, these icy pebbles can also potentially become trapped in the outer disk, beyond the snowline \citep[e.g.,][]{Pinilla12}. 
The mechanism responsible is trapping in the gas pressure maxima induced by the presence of forming planets that are also responsible for the creation of mm dust emission rings.

\begin{figure}\label{co_figure}
\includegraphics[width = \textwidth]{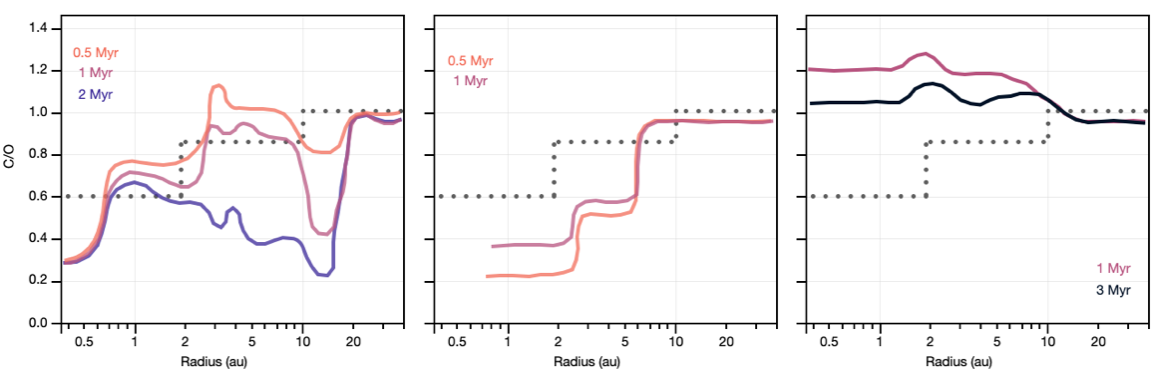}
\caption{Examples of how the C/O value in the disk midplane can vary in time with chemistry \citep[left;][]{Eistrup18}, pebble drift \citep[middle;][]{Booth17}, and pebble trapping \citep[using the models presented in][]{Booth19}. The dashed lines correspond to the C/O variations due solely to the positions of snowlines \citep{Oberg11}. 
Figure has been adapted from the original credited to L.~I.~Cleeves which used data from the cited works.}
\end{figure}

Figure~\ref{co_figure} shows examples from models of how C/O in the disk midplane can be affected by snowlines \citep[dotted lines in all panels;][]{Oberg11}, chemistry \cite[left panel;][]{Eistrup18}, icy pebble drift \citep[middle panel;][]{Booth17}, and pebble trapping beyond the snowline \citep[right panel;][]{Booth19}. 
Due to snowlines alone, and considering only the main oxygen and carbon carriers, the gas becomes progressively carbon rich due to the freeze-out of \ce{H2O} (at $\approx 2$~au) then \ce{CO2} (at $\approx 10$~au).  
Beyond $\sim 10$~au only CO remains in the gas-phase leading to C/O $\approx 1$ \citep{Oberg11}. 
Chemistry alone acts to decrease the C/O ratio over long timescales ($\gtrsim 1$~Myr) as less volatile species (e.g., \ce{CO2}) are converted by cosmic-ray-induced chemistry into more volatile species which remain in the gas-phase and enrich the gas in oxygen (e.g., \ce{O2}). 
This effect is sensitive to the assumed cosmic-ray ionisation rate in the disk midplane \citep{Eistrup18}.
Pebble drift allows oxygen-rich ices to be transported inwards, enriching the gas in oxygen and lowering the C/O ratios at the positions of key snowlines where key oxygen-bearing species sublimate \citep{Booth17}. 
If icy grains are trapped beyond the snowline in e.g., pressure traps invoked by forming planets, then only the gas (dominated by volatiles such as CO and \ce{CH4}) can be transported inwards, enriching the inner disk in carbon-bearing species and leading to  C/O $\gtrsim 1$. 
The model results in the right panel are generated using the models described in \citet[][]{Booth19} in which a Jupiter-mass planet has carved a gap at 20~au (R.~A.~Booth; priv.~comm.).

\subsection{Insights from ALMA studies of protoplanetary disks at high resolution}
\label{PPDsALMAhigh}

Moderate-resolution observations of molecular line emission from protoplanetary disks with ALMA have revealed many intriguing properties of the gas, including the aforementioned evidence of radial and vertical gradients in gas composition and metallicity; however, most results discussed thus far are probing emission on relatively large scales, usually at or beyond size scales similar to the size of the solar system ($\gtrsim$~50~au).
Whilst challenging, it is possible to observe molecular emission from protoplanetary disks at higher resolution, $\sim 0\farcs1$ or at a spatial scale of $\sim$~15~au for a disk at a distance of $\sim$~150~pc. 
This has been predominantly done for CO and its isotopologues, because these lines are bright, and usually with the aim to resolve structure and kinematic features in the disk gas.
However, this resolution has only been achieved in a handful of disks in other key molecular tracers. 

\citet{Nomura21}~presented high-resolution ($\approx 9$~au) observations of CO, CN, and CS in the disk around TW Hya. 
They confirmed that either oxygen or CO depletion was needed in the outer disk to explain the emission from CN and CS, in agreement with previous work.  
However, because they also resolved the inner region of the disk, they suggested that a high dust-to-gas mass ratio ($\sim 10$) was needed in the inner disk ($\lesssim 20$~au) to explain suppression of the emission from CN inside this region. 
This high dust-to-gas mass ratio is consistent with the hypothesis of radial drift of dust grains. 
Although not discussed at length in that paper, the emission profiles from all molecular tracers show radial variations that are not always coincident with features the dust thermal emission at the same resolution, although it should be noted that the dust sub-structure in TW Hya is more subtle than in other disks.  

\begin{figure}\label{MAPS_plots}
\includegraphics[width = \textwidth]{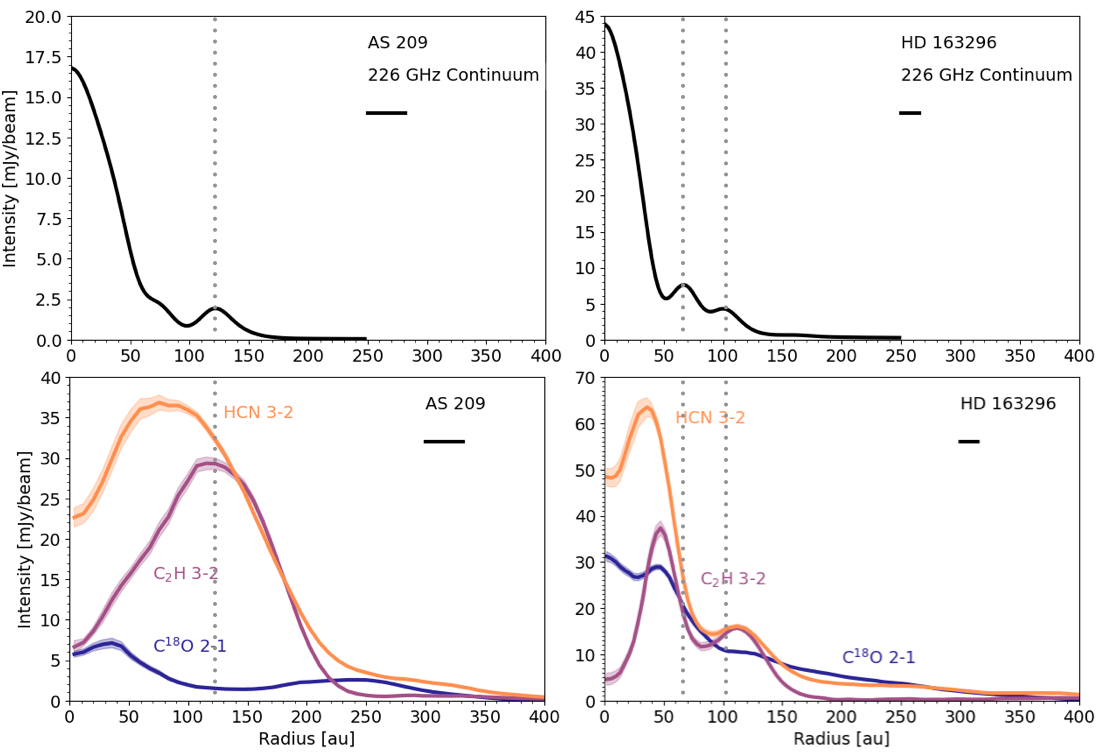}
\caption{Azimuthally-averaged radial profiles of the continuum emission at 226~GHz (top), and line emission from HCN, \ce{C2H}, and \ce{C^{18}O} (bottom) from the disks around AS~209 (left) and HD~163296 (right). 
All data are from the ALMA Large Program, MAPS \citep{Oberg21b,Law21a,Guzman21,Sierra21}.  
The horizontal bars indicate $0\farcs15$ which is the resolution of the data (coinciding to 32~au and 15~au in AS~209 and HD~163296, respectively.)
The vertical dotted lines indicate prominent continuum rings to highlight that it is not always the case that molecular emission rings coincide with those in the dust continuum. 
Note that the profiles presented here differ slightly from those presented in \citet{Guzman21} as we show the full azimuthally-averaged profiles, whereas presented in that work, are averages over a 30\degree~wedge along the disk major axis.}
\end{figure}

This property of highly-resolved line emission (i.e., that it is not always coincident with observed sub-structure in the dust) was also a key conclusion of the ALMA Large Program, MAPS (Molecules with ALMA on Planet-forming Scales; see \citealt{Oberg21b} for an overview of MAPS\footnote{\url{https://alma-maps.info/}}). 
This program surveyed five protoplanetary disks in a suite of key molecular tracers at high resolution (down to $\sim~0\farcs1$ for lines in ALMA Band 6).  
\citet{Law21a}~identified all instances of chemical sub-structure (gaps, rings, and plateaus of emission) and compared the identified sub-structures within and between families of molecules (e.g., the hydrocarbons versus the nitriles) and with any sub-structure in the dust at mm wavelengths.  
Similar to that found for TW Hya, there is not always a coincidence between, for example, a ring in dust emission and either a ring or a gap in molecular line emission. 
To demonstrate this, presented in Fig.~\ref{MAPS_plots} are the continuum emission at 226~GHz (top) and the azimuthally-averaged line emission from HCN, \ce{C2H}, and \ce{C^{18}O} (bottom) for the disks around AS~209 (left) and HD~163296 (right) at a resolution of $\approx 0\farcs15$.  
The dotted vertical lines highlight the prominent continuum rings in both disks.
A follow-up analysis by \citet{Jiang22}~confirmed no statistically significant evidence for a global correlation between structure in line emission and that in dust emission.  
All-in-all this indicates that there exist radial gradients in composition on even smaller scales that previously inferred, and that there is also complex coupling between the gas and dust evolution, the chemistry within the disk, and the emergent molecular line emission.

Another result of interest for planet formation from MAPS is that the data quality were high enough to directly trace the heights of the emitting surface in the molecular line channel maps \citep[][]{Law21b}.  
As expected, the CO line emission was found to arise at larger scale heights in the disk atmosphere; however, the line emission from \ce{HCN} and \ce{C2H} in the disks around AS~209, HD~163296, and MWC~480, was found to arise from deeper down in the disk, at $z/r$ close to or below 0.1, i.e., below a height of 5~au at a radius of 50~au.  
Recall that bright emission from \ce{C2H} is a strong indicator of carbon-rich gas.
To put these numbers in some context, a 5~M$_\mathrm{J}$ planet forming at 50~au around a Sun-like star has a Hill radius of $\approx$~6~au.  
The feeding zone of a gas-giant planet is thought to extend to a few Hill radii\footnote{The Hill radius for a planet in a circular orbit is given by $r_\mathrm{H} = a \left( \frac{\mathrm{M_p}}{\mathrm{3 M_\star}}\right)^{1/3}$ where $a$ is the planet's semi-major axis, and $\mathrm{M_p}$ and $\mathrm{M_\star}$ are the masses of the planet and star, respectively}; hence, for these three disks, at least, there is the potential for any forming planets therein to accrete carbon-rich gas.

A recent moderately-high resolution ($0\farcs2 - 0\farcs4$) chemical survey towards the disk, HD~100546, has revealed further proof of the potentially extreme radial gradients in composition that are possible in protoplanetary disks \citep[][]{Booth24}.
HD~100546 is a Herbig Ae star (T$_\mathrm{eff} \approx~10,000$~K) which is host to a large protoplanetary disk ($\approx 400$~au in radius measured in CO emission) in which 
the mm dust has been sculpted into two clear rings \citep[one centred at $\approx 25$ au and another at $\approx 200$ au; e.g., ][]{Fedele21}.  
Unlike the MAPS sources discussed above, this is one example of a disk which does appear to have coincident dust and chemical rings.  
There is an almost complete absence of emission from all molecules between the dust rings, with the exception of that from CO and \ce{HCO+}.  
The proposed tracers of carbon-rich gas (\ce{HCN}, \ce{CN}, \ce{C2H}, and \ce{CS}) all possess a bright emission ring in the outer disk, which is mostly coincident with the outer dust ring.
Further there is evidence of a gradient in gas-phase C/O across the inner dust ring: emission from O-bearing species (e.g., SO, \ce{SO2}, and \ce{CH3OH}) peak in emission just inside the inner dust ring ($\lesssim 10$~au), whereas that from hydrocarbons and nitriles peaks towards the outer edge of the inner dust ring ($\approx 30 - 50$ au). 
This disk hosts both an O-rich gas-phase reservoir (revealed by ice sublimation in the inner disk) and a C-rich reservoir which itself varies in abundance across the disk. 
This strong association between molecular emission rings and the dust rings, make this disk an interesting test case for the coupling between dust evolution and chemistry.

\subsection{Recent results from JWST}
\label{PPDsJWST}

Whilst resolved observations with ALMA have revealed a wealth of information on the composition, and presence of chemical gradients, on size scales from $\approx 10 - 50$~au, these data are not yet able to probe the composition of the very innermost disk region, that overlaps with orbital locations of the population of gas giants detected via the transit method and that are ripe for atmospheric characterisation.  
To probe this hot molecular reservoir, we need to move to infrared (IR) wavelengths, and in particular, mid-IR wavelengths, where many key carriers of oxygen, carbon, sulphur, and nitrogen (e.g., \ce{CO2}, \ce{C2H2}, \ce{HCN}, \ce{NH3}, and \ce{SO2}) have strong infrared transitions \citep[e.g.,][]{Bast13}. 

Early observations of protoplanetary disks with the \emph{Spitzer} Space Telescope at mid-IR wavelengths revealed some interesting trends.  
There appeared to be a dearth of \ce{H2O} observed in disks towards higher-mass stars compared with lower-mass stars, although OH was regularly detected towards the former \citep[e.g.,][]{Pontoppidan10}. 
There appeared a trend in \ce{C2H2}/HCN ratio with decreasing stellar mass, with later-type stars having stronger \ce{C2H2} emission compared with HCN emission than Sun-like stars \citep[e.g.,][]{Pascucci09}. 
Finally, there was noted a trend in the ratio of HCN-to-\ce{H2O} mid-IR line emission with respect to disk mass (as measured from mm wavelength observations), with more massive disks having stronger HCN emission relative to that for \ce{H2O} 
\citep[e.g.,][]{Najita13}.  
Several possible explanations for these trends have been proposed.  
The dearth of \ce{H2O} detections towards Herbig disks may be related to the strong UV coupled with the presence of gas cavities typical of these sources \citep[e.g.,][]{Banzatti17}.
The increase in \ce{C2H2}/HCN emission with decreasing stellar mass may be related to the higher X-ray luminosities relative to their UV luminosity of later-type stars compared with Sun-like stars.  X-rays are able to break apart CO more effectively than UV photons releasing free carbon and driving a rich ion-molecule chemistry to produce hydrocarbons \citep[e.g.,][]{Walsh15}.
And finally, the trend in HCN/\ce{H2O} may be because more massive disks are able to lock away a larger fraction of their material in icy planetesimals beyond the snowline, such that the inner disk in these systems are fed with O-poor gas \citep[e.g.,][]{Banzatti20,Kalyaan23}. 

By re-opening access to the mid-IR, JWST now allows some further testing of these interesting trends seen in the \emph{Spitzer} sample.  
However, so far, most reported results are focused on single objects rather than population level studies, the analysis of which will likely come later in the various observing programs. 
Here, I give a brief overview only of what has been found so far:  
see \citet[][]{Pontoppidan24} and \citet[][]{Henning24} for overviews of the JDISC (JWST Disk Infrared Spectral Chemistry Survey) and MINDS (MIRI mid-INfrared Disk Survey) collaborations, respectively.  
An overview of MINDS early results can also be found in \citet[][]{vanDishoeck23} and 
\citet[][]{Kamp23}.

Early results from JWST spectroscopy is revealing that the chemical composition of the inner regions of disks is as diverse as that found for the outer disk at longer wavelengths. 
The increase in sensitivity provided by JWST allowed the first detection of \ce{^{13}CO2} in the disk around GW Lup (an M1.5 star in the Lupus star-forming region), which was identified to be a \ce{CO2}-rich disk in observations with \emph{Spitzer} \citep{Grant23}. 
Fitting of the data confirmed that the \ce{CO2} is emission is optically thick, with a higher excitation temperature than \ce{^{13}CO2} (400~K versus 325~K).
Emission from OH, \ce{H2O}, \ce{C2H2}, and \ce{HCN} were also confirmed, although the \ce{H2O} was found to be relatively weak, with an inferred ratio of the \ce{CO2}-to-\ce{H2O} column density of 0.7.  
Such a high ratio is unexpected, but can be explained by inner disk structure: a gas cavity in the disk for which the outer edge lies between the \ce{H2O} and \ce{CO2} snowlines, can explain the weak water emission with respect to that for \ce{CO2} \citep{Vlasblom24}. 
This result demonstrates how spatially unresolved mid-IR spectroscopy can reveal inner disk structure.

JWST observations have also suggested that compact disks around T Tauri stars may be more water rich than more radially extended and structured disks, a result that might be expected due to more massive and extended disks already hosting planets, that have trapped a significant fraction of water in ice form beyond the water snowlines \citep[as discussed previously;][]{Banzatti23}. 
More compact disks have undergone more extreme dust drift indicating a lack of planets (or indeed other means of trapping) that would otherwise halt pebble drift. 
This survey of four disks (two compact and two more extended) with JWST/MIRI showed that, whilst all four sources possessed a hot water component in their spectra, only the two compact disks possessed an excess cooler water component with a representative excitation temperature of $170 - 400$~K with a characteristic emitting radius within $\sim 1$~au. 
This was explained by compact disks possessing a higher inward flux of icy pebbles creating a substantial reservoir of cool water at and around the position of the water snowline.  
This confirms, again, that disks can possess composition gradients due to the coupling between dust evolution and chemistry.

JWST spectroscopy of T~Tauri disks so far confirm the \emph{Spitzer} results, that water is apparently ubiquitous in such systems. 
The MINDS survey has so far confirmed the presence of \ce{OH}, \ce{H2O}, \ce{CO}, \ce{HCN}, and \ce{CO2}, in the disks around SY Cha and Sz~98: both disks exhibit inner disk chemistry consistent with O-rich gas \citep[][]{Gasman23,Schwarz24}. 
Very recent results for the nearest T Tauri disk, TW Hya \citep{Henning24}, also confirm the presence of \ce{OH}, \ce{H2O}, and \ce{CO2} with non-detections reported for HCN, \ce{C2H2}, \ce{CH3}, and \ce{CH4}.  
This again points to O-rich gas in the inner regions of TW Hya, although a depletion factor for O, N, and C of $\approx 50$ within 2.4~au was needed to better reproduce the spectrum.
Water and \ce{CO2} have also been detected in the inner disk of PDS~70, the only protoplanetary disk, so far, to host multiple protoplanets \citep[][]{Perotti23}. 
This is despite PDS~70 hosting a large dust cavity within which there are two confirmed protoplanets \citep[and a potential third planet revealed in JWST high-contrast imaging;][]{Christiaens24}.
The disk around AS~209 (also a MAPS source) is also found to be host to hundreds of water lines, with only marginal detections of \ce{CO2} and HCN \citep{MunozRomero24}.  
As a highly-structured disk, this source was also found to be host to a tentative cool water component: this sits somewhat in contrast with the findings discussed above, that only compact disks should be host to this cool component. 
Despite this being a tentative result, it hints that some structured disks do have some degree of icy pebble drift enhancing the water abundance at the snowline.
It should be noted here that both SY~Cha and AS~209 may be infrared-variable sources, with differences in line fluxes reported for the current JWST era compared with previous observations with \emph{Spitzer} \citep{Schwarz24,MunozRomero24}. 

What is interesting about this set of results for T Tauri disks, is that ALMA has revealed that most of them (with the exception of SY Cha) have indicators of carbon-rich outer disks, including the outer disk of the protoplanet-hosting PDS~70 \citep{Facchini21}.  
So there appears to be emerging a pattern (of course to be confirmed) that the inner regions of disks around T Tauri stars are host to O-rich gas, but with some indicators that disk structure on small scales is influencing the relative fluxes and strengths of the lines.
So what about the disks around lower-mass stars? 
So far, two JWST spectra have been published for disks around cooler M~Dwarf star that show a potentially more diverse picture.  
The disk around the very low-mass star (0.14 M$_\odot$), J160532, exhibits a very different chemistry: confirmed in this disk are the first detections of \ce{C6H6} and \ce{C4H2} as well as other hydrocarbons such as \ce{C2H2} and, tentatively, \ce{CH4} \citep{Tabone23}. 
Whilst \ce{CO2} is detected, an upper limit only for \ce{H2O} was obtained.  
All-in-all this disk exhibits a very hydrocarbon rich inner disk indicative of carbon-rich gas (C/O $>$ 1).

In contrast, the disk around the M5 star (0.17 M$_\odot$), Sz~114, is host to a wealth of \ce{H2O} lines, in hand with detections of CO, \ce{CO2}, HCN, and \ce{C2H2} \citep{Xie23}.  
Hence, this source appears to host oxygen-rich inner disk gas more in line with that seen around higher-mass T~Tauri stars.  
The authors speculate that the young nature of this source (compared with J160532) coupled with its particularly massive and extended disk, may have helped to preserve a water rich inner disk.  
However, again, this result sits in contrast with the idea that structured disks should not be host to a cooler water component which is indicative of icy pebble sublimation around the water snowline \citep[similar to that found for AS~209;][]{MunozRomero24}. 
It is possible that for some sources, the structures are too weak to fully impede the inwards drift of icy pebbles, and this should certainly be tested with models.
As the most common type of star in the galaxy, quantification of the diversity in composition across the M-Dwarf disk population is of great importance for understanding what is likely to be the most common planetary system host.

\section{What have we learned about the composition of gas-giant exoplanet atmospheres?}
\label{Exoplanets}

\subsection{Insights from the pre-JWST era}
\label{exopreJWST}

We have learned from both ALMA and JWST that disks are host to sometimes extreme vertical and radial compositional gradients, that is likely to influence the composition of planets forming therein.  
That the population of gas-giant planets may have a fingerprint of their formation in their atmosphere, comes from the idea that disks possess relatively sharp transitions in C/O (and metallicity) in their midplanes due to the unique locations of key snowlines including \ce{H2O}, \ce{CO2}, and \ce{CO} \citep{Oberg11}.  
In highly structured disks, that ironically are likely to have been shaped by ongoing planet formation, we can see localised radial variations in C/O ratio that is not due only to snowlines, but is likely also due to coupling between dust and gas evolution with chemistry. 
We have also seen that for the same system (e.g., TW Hya) the inner disk gas can exhibit a different C/O ratio to that in the outer disk, with inner disks having more oxygen-rich gas than their outer disks \citep[e.g.,][]{Bergin16,Henning24}.
This general trend does align with the concept of snowlines, as in general, we expect that disk midplane gas should become progressively more carbon-rich as first \ce{H2O} and then \ce{CO2} freeze out (see Fig.~\ref{co_figure}).
Coupled with this, the population of hot gas giants, in particular, whilst ripe for atmospheric characterisation, may have undergone a variety of processes that will have shaped their atmospheric composition, including migration through the disk, potentially through and across snowlines, and atmospheric pollution by accreting pebbles and planetesimals \citep[see recent work by, e.g.,][]{Kessler23}.  
Of course, this is assuming that the planet originally formed in the outer disk via core accretion, similar to the gas giants in the solar system.
Nonetheless, there still persists the idea that snowlines are a key influence in the composition of the atmosphere, thus revealing (potentially) the original formation location and formation mechanism of the planet.  
Hence, derivation of the C/O ratio and metallicity of the atmosphere are key parameters that are extracted from exoplanet spectra (both in transmission and emission).

What have these results revealed so far?  
Conveniently, recently published is a compilation of the pre-JWST C/O ratios derived for both transiting and directly-imaged gas-giant planets \citep{Hoch23}. 
This is a useful dataset with which to compare more recent results with JWST.  
The C/O retrievals for the transiting planets presented in \citet{Hoch23} were all performed in a systematic way on eclipse data, with 17 of the 25 exoplanets also possessing transmission spectra \citep{Changeat22}.  
We also show here the suite of retrievals performed on pre-JWST transmission spectra of 16 gas giants using the ARCiS spectral retrieval code as a comparison using a different tool \citep{Kawashima21}. 
We include their results for both equilibrium and disequilibrium chemistry to show the variation in C/O retrieved by introduction of this parameter, which is another complicating feature in the interpretation of spectroscopy of exoplanetary atmospheres. 

\begin{figure}[h]
    \includegraphics[width=\textwidth]{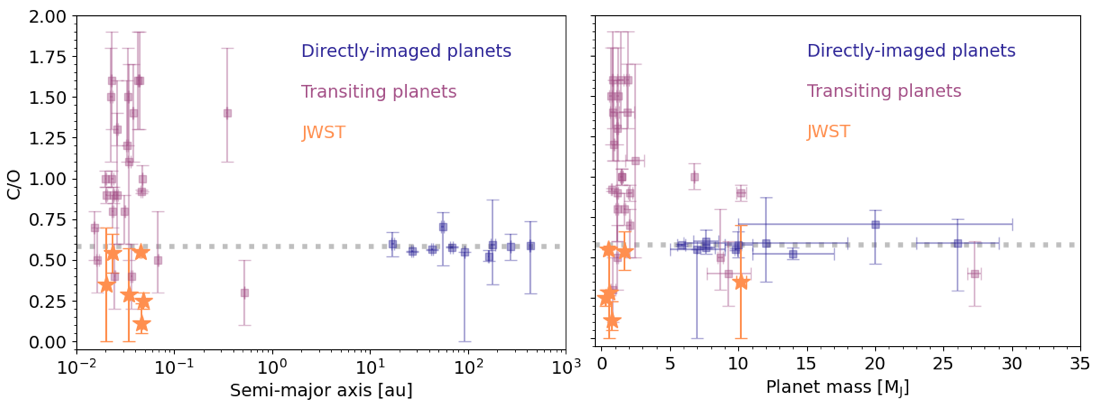}
    \caption{Global atmospheric C/O values for directly-imaged (blue) and transiting (purple) planets as a function of semi-major axis (left) and planet mass (right). This plot is reproduced using the data compiled in \citet{Hoch23}. Added to this plot are values from current C/O values reported in the literature \citep{Ahrer23,Alderson23,Feinstein23,Rustamkulov23,Tsai23,Edwards24,Smith24,Xue24,Bell23,Radica23,Taylor23,Coulombe23} from the analysis (or inclusion) of data from JWST (orange).  Error bars are included only if reported. 
    The dotted horizontal line shows the solar C/O value \citep[0.58;][]{Asplund21}.}
    \label{exo-CO-plot}
\end{figure}

\begin{figure}
    \includegraphics[width=\textwidth]{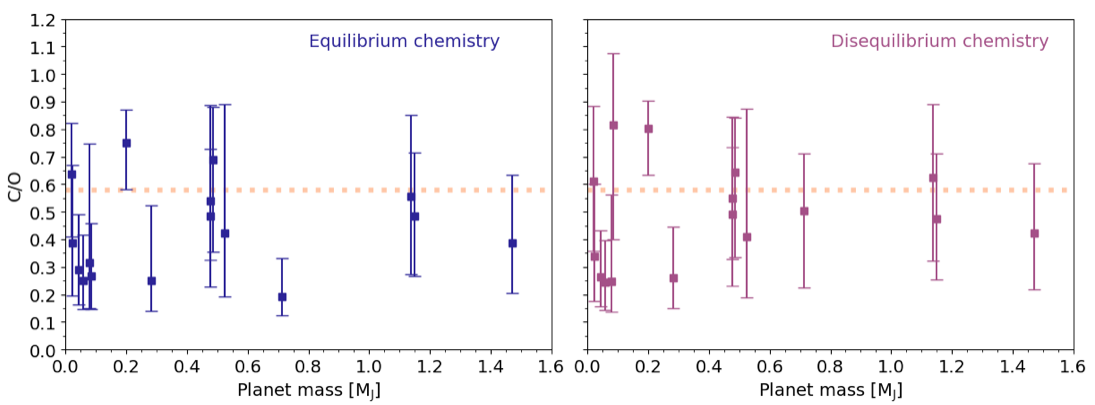}
    \caption{Global atmospheric C/O values for transiting planets using the ARCiS retreival code assuming equilibrium chemistry (left) and disequilibrium chemistry (right). 
    Figure is produced from data and scripts from \citet{Kawashima21}. 
    The dotted horizontal line shows the solar C/O value \citep[0.58;][]{Asplund21}.}
    \label{arcis-CO-plot}
\end{figure}

The population of directly-imaged planets presented in \citet{Hoch23} all exhibit C/O values ranging from 0.52 to 0.704, i.e., all have oxygen-rich atmospheres (see blue points in Fig.~\ref{exo-CO-plot}). 
As mentioned previously, the population of directly-imaged planets are younger ($\lesssim 200$~Myr), typically more massive ($\gtrsim 6$~M$_\mathrm{J}$), and tend to be found around higher-mass (i.e., A-type) stars, than the population of transiting planets that have had their atmospheres characterised 
($600~\mathrm{Myr} \lesssim A_\mathrm{p} \lesssim 7,000~\mathrm{Myr}$, and $0.7~\mathrm{M_{J}} \lesssim M_\mathrm{p} \lesssim 27~\mathrm{M_{J}}$).  
The population of transiting planets (purple points in Fig.~\ref{exo-CO-plot}) have C/O values derived from eclipse (usually combined with transmission) spectra ranging from 0.3 to 1.6; hence, within the transiting exoplanet population there is both a wider dynamic range in C/O, a population that appears to be dominated by exoplanets with super-solar ($> 0.58$) C/O ratios (80\%), and a substantial fraction of which have C/O $> 1$ (52\%). 
In contrast, the retrievals on transmission spectra only presented in \citet{Kawashima21} show a narrower distribution with all retrievals predicting oxygen-rich gas (C/O $< 1$; see Fig.~\ref{arcis-CO-plot}): assuming equilibrium chemistry (blue points in left panel) the C/O range is from 0.19 to 0.75, whilst that assuming disequilibrium chemistry (purple points in right panel) is from 0.25 to 0.82.  
On average, the models with disequilibrium chemistry predict a slightly more carbon-rich gas than those assuming equilibrium chemistry. 
These retrievals also predict only 19\% (31\% assuming disequilibrium chemistry) of transiting planets having super-solar C/O with 100\% having C/O~$< 1$.
It should be pointed out here that the two retrieval studies discussed here are fitting overlapping datasets.  
The challenges in retrievals of atmospheric parameters from exoplanetary spectra are well described and discussed in \citet{Barstow20}, with the results being very dependent on the underlying assumptions in the models (i.e., an isothermal atmosphere versus one with a temperature-pressure profile).   

So what does this tell us about the nature of the planets?  
That the directly-imaged population have both a narrower range in C/O and one that is consistent with solar hints that this population of planets may have formed via gravitational instability. This is because, having collapsed directly from the disk material, their composition should be reflective of the global metallicity of the parent cloud and host star.   
Of course, planets formed by this mechanism could be enriched post formation via the accretion of planetesimals, but this is likely to have a relatively minor effect on the atmospheric composition, given the already large mass of the planet at the point of formation.
Indeed, that this population of exoplanets could be distinct from the population of transiting planets is mentioned in \citet{Hoch23}. 
What is also particularly intriguing is that this population of exoplanets also overlaps, in terms of semi-major axis, with the indirect features of ongoing planet formation seen in protoplanetary disks, in the form of gas and dust gaps, and kinematic perturbations of the gas due to a hidden accreting planet \citep[see Fig.~14 in][]{Pinte23}.  
Recall that these regions of protoplanetary disks have been shown to be composed of carbon-rich and metal-poor gas; hence, if these planets are actively accreting from this gas, we would expect them to be enhanced in carbon, and depleted in heavy elements. 
That we don't see signatures of this in the population of directly-imaged planets, is also evidence of likely early formation via gravitational instability.  
This is consistent with this particular class of disk possessing rather extreme structures and chemistry: early planet formation coupled with dust and gas evolution has shaped the disk into what we see today.
What remains a puzzle, is that attempts to \emph{directly} detect these embedded protoplanets suggested in ALMA data have only provided upper limits, with the exception of PDS~70, which remains the only concrete detection confirmed with multiple facilities and techniques \citep[e.g.,][]{Keppler18,Haffert19}.  
Indeed, upper limits on planet mass inferred from high-contrast imaging suggest planet masses no higher than a few Jupiter masses \citep[e.g.,][]{AsensioTorres21}, which is in contrast with the mass range of directly-imaged planets.  
However, aside from the presence of gaps, there are \emph{indirect} signatures of embedded accreting planets in disks through kinematic perturbations of the background disk gas that is visible in channel maps of molecular emission observed with ALMA \citep{Pinte23}. 

As for the transiting planets, the wider dynamic range in C/O suggests more diverse formation pathways and outcomes consistent with core accretion and migration \citep[e.g.,][]{AliDib17}, although forming a hot Jupiter with a carbon-rich atmosphere usually requires some additional source of carbon in the gas phase such as destruction of carbon grains \citep[e.g.,][]{Cridland19}. 
Indeed this concept of the presence of a ``soot line" in protoplanetary disks and its evolution is now being discussed in the context of how planets that assemble from solid material, like the Earth, are carbon poor \citep[e.g.,][]{Li21}; but, of course, another consequence of this is that there can exist a carbon-rich gas-reservoir within this defined region of the disk, within which gas giants would have the opportunity to accrete carbon-rich gas.

\subsection{Recent results from JWST}
\label{exonowJWST}

So how do the more recent results from JWST compare with the results from the pre-JWST era?  
JWST re-opens the window in the mid-IR and also offers higher sensitivity and spectral resolution at near-IR compared with \emph{Spitzer}.  
As noted by the retrievals presented in \citet{Changeat22}, because the carbon-bearing molecules are not as regularly detected as \ce{H2O} there remain large error bars on the C/O ratios derived from previous generation data.
Here I focus on the population of hot gas giants observed with JWST as these, again, are most likely to have retained some chemical fingerprint of their formation. 

The first spectrum published was the early release science (ERS) data for the hot Saturn, WASP~39b (0.3~M$_\mathrm{J}$), which was observed with four JWST instruments/modes: NIRSPEC/G395, NIRSPEC/PRISM, NIRISS, and NIRCam \citep{Ahrer23, Alderson23, Feinstein23, Rustamkulov23}. 
Independent fitting of the spectra with each instrument/mode suggests a high metallicity atmosphere (of order 10 times solar) and a sub-solar C/O ratio ($\sim 0.2 - 0.3$).  
In this data was also the first detection of \ce{SO2} in an exoplanet atmosphere, that also confirms its oxygen-rich high-metallicity nature \citep{Tsai23}. 
In contrast to WASP~39b, a combined analysis of data from HST and Gemini South/INGRINS with that from JWST/NIRSPEC for WASP~77Ab (1.7~M$_\mathrm{J}$), suggests a solar C/O \citep[$0.54\pm0.12$;][]{Edwards24} but a metal-depleted atmosphere with an inferred metallicity of [(C+O)/H]~$=-0.61_{-0.09}^{+0.10}$ \citep{Smith24}.  
This is consistent with WASP~77Ab having formed beyond the water snowline in its natal protoplanetary disk \citep{Reggiani22}.
Observations of the canonical hot Jupiter, HD~209548b (0.7 M$_\mathrm{J}$) with JWST/NIRCam transmission spectroscopy suggested a low C/O ratio of 0.11 with an metallicity of around 3 times solar \citep{Xue24}.  
JWST/NIRCam transmission and emission spectroscopy of WASP~80b (0.6~M$_\mathrm{J}$) revealed concrete evidence for the presence of \ce{CH4} in the atmosphere of an exoplanet suggesting a C/O value less than solar ($< 0.57$) and that is metal enriched \citep[around 5 times solar;][]{Bell23}. 
The JWST/NIRISS transit spectrum of WASP~96b (0.5~M$_\mathrm{J}$) is reproduced with a solar C/O ratio and a solar-to-super-solar metallicity \citep{Radica23,Taylor23}.
Using a different type of observation, thermal emission from the dayside of WASP~18b (10 M$_\mathrm{J}$) was observed with JWST/NIRISS, analysis of which suggests a solar metallicity and an upper limit to C/O of $< 0.7$ determined using various retreival methods and assumptions in the underlying models \citep{Coulombe23}. 

Figure~\ref{exo-CO-plot} shows a compilation of C/O values for directly-imaged (blue) and transiting planets (purple) from the pre-JWST era as compiled by \citet{Hoch23}. 
Added to this plot are those values already discussed that include or analyse data from JWST (orange stars).  
Whilst it is early in the lifetime of the telescope, so far, data from JWST is suggesting that this population of hot gas-giant planets are either consistent with the solar value or  oxygen-rich (with C/O $< 1$), with some extremely enriched in oxygen (e.g., HD~209458b). 
The population is tentatively showing a wider diversity in C/O than the directly-imaged population.
It is worth to note that the JWST data have revised the C/O value for HD~209458b from 0.92 to 0.11 \citep{Changeat22,Xue24}, and that for WASP~77Ab from 0.8 to 0.54 \citep{Changeat22,Edwards24}, showcasing the power of JWST data to more tightly constrain the composition of the atmospheres of this class of planets.
Hence, as yet, there is no confirmation using data from JWST for the existence of a gas giant exoplanet with a carbon-rich atmosphere (C/O~$> 1$).  
 
\section{Final remarks and a look forward}
\label{Conclusions}

\begin{figure}
    \centering
    \includegraphics[width=0.6\textwidth]{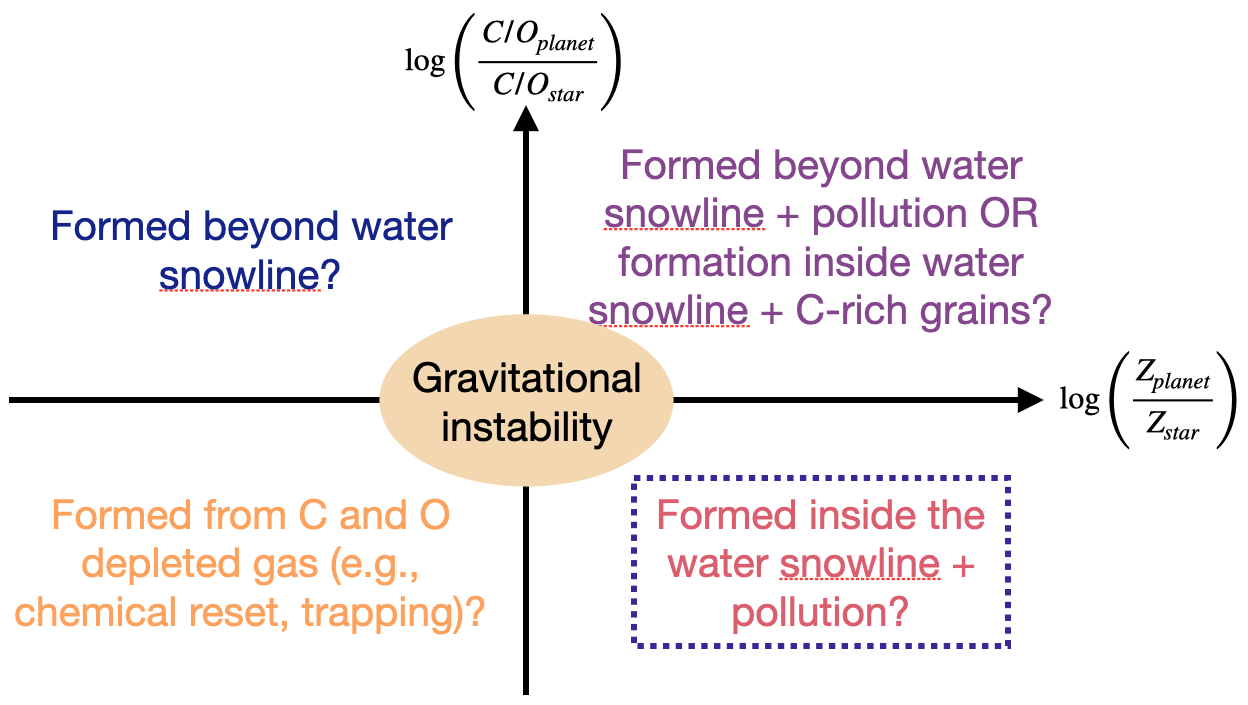}
    \caption{A range of possible interpretations of the C/O and metallicity measured in a gas-giant planet atmosphere.  
    The $x$ axis represents metallicity and the $y$ axis represents the C/O value, such that the left and right of the diagram represented sub-stellar, and super-stellar metallicity, respectively, and the top and bottom of the diagram represent super-stellar and sub-stellar C/O values, respectively. 
    Figure reproduced from an original by J.~Teske based on the discussion in \citet{Reggiani22}.}
    \label{co-interpretation-diagram}
\end{figure}

Discussed in this short review is an overview of the current understanding on the chemical composition and diversity of the planet-forming regions of protoplanetary disks contrasted with that present in the gas-giant planet population.  
It is clear that the gas in nearby protoplanetary disks can exhibit sometimes extreme compositional gradients in C/O and metallicity in both the radial and vertical directions that are expected to influence the composition of actively-accreting planets forming therein.  
These gradients can be created by a combination of dust evolution and chemistry; however, high-resolution observations with ALMA suggest that there is not always a one-to-one correlation between mm continuum rings and molecular emission rings, hinting at a complex interplay between small-scale disk structure, chemistry, and emergent line emission.  
What these small-scale gradients mean for planet formation is not yet clear. 
Further, there is now compelling evidence that the phenomenon of icy pebble drift is influencing the composition of the inner regions of protoplanetary disks creating an O-rich inner disk in conjunction with an C-rich outer disk.  

These compositional trends in protoplanetary disks can now be contrasted with those in the exoplanet population.  
The population of directly-imaged exoplanets, which overlap in orbital location with the presence of rings and gaps in protoplanetary disks, appear to have atmospheres that are O-rich and consistent with solar metallicity.  
That this region of disks tends to be metal-depleted and O-poor, suggests that this population of planets may have formed early in the protoplanetary disk lifetime, before the effects of dust evolution and chemistry have taken hold. 
The timescale for planet formation via core accretion ($\approx 10$~Myr) is longer than that for the discussed depletion factors (typically a few Myr), although pebble accretion can significantly speed up core formation \citep[e.g.,][]{Bitsch15}. 
An alternative explanation is that these planets formed early in the disk via gravitational instability which would explain why their atmospheric metallicity is so far consistent with the solar value.
In tandem, JWST is providing now much tighter constraints on the atmospheric composition of the population of hot gas giants.  
The results of retrievals performed on pre-JWST data suggested a wide dynamic range in atmospheric C/O values with a significant population of planets with super-solar C/O, although this conclusion is dependent on the retrieval model and underlying assumptions.  
Wrapping in the higher sensitivity and spectral resolution data from JWST has started to revise this picture.  
All characterised planets, so far, are consistent solar or sub-solar C/O, and are typically metal enriched, with the exception of WASP~77Ab which is metal depleted \citep{Edwards24,Smith24}.
It should be noted that this is based on results for six planets only and such a broad conclusion could be revised when more results from JWST are published.

So what does all this mean for our understanding of planet formation? 
In Fig.~\ref{co-interpretation-diagram} I show a schematic of possible interpretations related to the origin of a gas-giant planet based on measurement of its metallicity ($x$ axis) and C/O ($y$ axis) relative to the stellar value.  
A planet that is depleted in metallicity with a super-stellar C/O could have formed beyond the water snowline (top left quadrant); a planet depleted in metallicity and with a sub-stellar C/O could have formed by C- and O-depleted gas consistent with some degree of icy pebble trapping beyond the water snowline (bottom left quadrant); one enhanced in metallicity and with a super-stellar C/O ratio could have multiple interpretations, e.g., it formed beyond the water snowline and has been polluted by planetesimals/pebbles rich in carbon, or it formed within the water snowline and accreted C-rich grains (top right quadrant); and finally, a planet enhanced in metallicity and with a sub-stellar C/O could have formed inside the water ice line and accreted heavy elements via planetesimals and/or pebble accretion (bottom right quadrant). 
So far, 5/6 gas giants characterised using JWST data are consistent with the interpretation presented in the bottom right quadrant as all are consistent with close-to solar or sub-solar C/O in conjunction with solar or super-solar metallicity (highlighted in the dashed box). 

So what do we need in the future to better confirm these tentative trends?  
More data on more objects, of course, is needed.  
With ALMA, there is a need for both population-level statistical studies to determine the compositional diversity of planet-forming disks and how this is affected by environment and age, as well as complementary high-resolution and high-sensitivity studies of some well-selected objects to enable a better understanding of the origin of small-scale chemical structure and its impact (if any) on planet formation. 
The planned ALMA Wideband Sensitivity Upgrade\footnote{\url{https://science.nrao.edu/facilities/alma/science_sustainability/wideband-sensitivity-upgrade}} will increase the available bandwidth in a single observation from 8~GHz to 32~GHz and at a spectral resolution suitable for astrochemistry ($\sim 0.1$~km/s). 
This will allow the detection of many more spectral lines in a single observation than currently possible, increasing the efficiency of the telescope by up to factor of $\approx 5$. 
JWST is already providing transformational insight into the composition of the inner regions of protoplanetary disks (mostly O-rich) and exoplanets (also mostly O-rich).  
Similar to that for disks, we need population-level statistical studies of the composition of gas-giant exoplanets.  
This will be possible in the near future with the Ariel mission, due for launch in 2026, and which will survey 1,000 exoplanets across a wavelength range covering spectral features of e.g., \ce{H2O}, \ce{CO2} and \ce{CH4} \citep[$1.25 -7.8~\mathrm{\mu m}$; ][]{Tinetti18}. 

So far, the tentative picture emerging is that the population of close-in gas-giant planets mostly have atmospheres consistent with formation at or within the water snowline, and  metal-enriched atmospheres consistent with pollution by accreting icy planetesimals and/or pebbles.  
Also, the population of far-out gas-giant planets that can be characterised by direct imaging, appear to be consistent with solar metallicity and C/O indicating possible formation via gravitational instability. 
It will be interesting to see in the coming years whether or not these arguably ``conventional" pictures of planet formation for both populations of planets are confirmed.

\end{document}